\documentclass[fleqn,usenatbib]{mnras}

\usepackage[T1]{fontenc}
\usepackage{ae,aecompl}
\usepackage{hyperref}

\usepackage{graphicx}	
\usepackage{amsmath}	
\usepackage{amssymb}	

\newcommand{\be}{\begin{equation}}
	\newcommand{\ee}{\end{equation}}
\newcommand{\ba}{\begin{eqnarray}}
	\newcommand{\ea}{\end{eqnarray}}
\newcommand{\brr}{\begin{array}}
	\newcommand{\err}{\end{array}}
\newcommand{\bc}{\begin{center}}
	\newcommand{\ec}{\end{center}}
\newcommand{\hm}{\,h^{-1}{\rm Mpc}}
\newcommand{\hk}{\,h^{-1}{\rm kpc}}
\newcommand{\msun}{\,h^{-2} {\rm M_\odot}}

\newcommand*{\swap}[2]{#2#1}

\title[Galaxy halo mass in the cosmic web environment]{Dependence of GAMA galaxy halo masses on the cosmic web environment from 100 square degrees of KiDS weak lensing data}
\author[M. M. Brouwer et al.]{Margot M. Brouwer$^{1}$\thanks{E-mail:brouwer@strw.leidenuniv.nl}, 
Marcello Cacciato$^{1}$, 
Andrej Dvornik$^{1}$, 
Lizzie Eardley$^{2}$, \and
Catherine Heymans$^{2}$, 
Henk Hoekstra$^{1}$, 
Konrad Kuijken$^{1}$, 
Tamsyn McNaught-Roberts$^{3}$, \and
Crist\'obal Sif\'on$^{1}$, 
Massimo Viola$^{1}$, 
Mehmet Alpaslan$^{4}$, 
Maciej Bilicki$^{1}$, \and
Joss Bland-Hawthorn $^{5}$, 
Sarah Brough$^{6}$, 
Ami Choi$^{2}$, 
Simon P. Driver$^{7, 8}$, \and
Thomas Erben$^{9}$, 
Aniello Grado$^{10}$, 
Hendrik Hildebrandt$^{9}$, 
Benne W. Holwerda$^{1}$, \and
Andrew M. Hopkins $^{6}$, 
Jelte T. A. de Jong$^{1}$,
Jochen Liske$^{11}$, 
John McFarland$^{12}$, \and
Reiko Nakajima$^{9}$,
Nicola R. Napolitano$^{10}$
Peder Norberg$^{3}$, 
John A. Peacock$^{2}$, \and
Mario Radovich$^{10}$, 
Aaron S. G. Robotham$^{7}$, 
Peter Schneider$^{9}$,
Gert Sikkema$^{12}$, \and
Edo van Uitert$^{13}$, 
Gijs Verdoes Kleijn$^{12}$
\\
\\
$^{1}$Leiden Observatory, Leiden University, Niels Bohrweg 2, 2333 CA Leiden, The Netherlands.\\
$^{2}$SUPA, Institute for Astronomy, University of Edinburgh, Royal Observatory, Blackford Hill, Edinburgh, EH9 3HJ, UK.\\
$^{3}$ICC \& CEA, Department of Physics, Durham University, South Road, Durham DH1 3LE, UK. \\
$^{4}$NASA Ames Research Centre, N232, Moffett Field, Mountain View, CA 94035, United States.\\
$^{5}$Sydney Institute for Astronomy, School of Physics A28, University of Sydney, NSW 2006, Australia. \\
$^{6}$Australian Astronomical Observatory, P.O. Box 915, North Ryde, NSW, Australia.\\
$^{7}$ICRAR M468, University of Western Australia, 35 Stirling Hwy, Crawley WA 6009 Australia.\\
$^{8}$SUPA, School of Physics and Astronomy, University of St Andrews, North Haugh, St Andrews, KY16 9SS, UK.\\
$^{9}$Argelander-Institut f{\"u}r Astronomie, Auf dem H{\"u}gel 71, D-53121 Bonn, Germany.\\
$^{10}$INAF-Osservatorio Astronomico di Capodimonte, Via Moiariello 16, 80131 Napoli, Italy. \\
$^{11}$Hamburger Sternwarte, Universit\"at Hamburg, Gojenbergsweg 112, 21029 Hamburg, Germany. \\
$^{12}$Kapteyn Astronomical Institute, University of Groningen, P.O. Box 800, 9700 AV Groningen, The Netherlands.\\
$^{13}$University College London, Gower Street, London WC1E 6BT, UK. \\
}

\date{Accepted XXX. Received YYY; in original form ZZZ}

\pubyear{2016}

\begin{document}
\label{firstpage}
\pagerange{\pageref{firstpage}--\pageref{lastpage}}
\maketitle

\begin{abstract}
Galaxies and their dark matter haloes are part of a complex network of mass structures, collectively called the cosmic web. Using the tidal tensor prescription these structures can be classified into four cosmic environments: voids, sheets, filaments and knots. As the cosmic web may influence the formation and evolution of dark matter haloes and the galaxies they host, we aim to study the effect of these cosmic environments on the average mass of galactic haloes. To this end we measure the galaxy-galaxy lensing profile of $91,195$ galaxies, within $0.039 < z < 0.263$, from the spectroscopic Galaxy And Mass Assembly (GAMA) survey, using $\sim100 \deg^2$ of overlapping data from the Kilo-Degree Survey (KiDS). In each of the four cosmic environments we model the contributions from group centrals, satellites and neighbouring groups to the stacked galaxy-galaxy lensing profiles. After correcting the lens samples for differences in the stellar mass distribution, we find no dependence of the average halo mass of central galaxies on their cosmic environment. We do find a significant increase in the average contribution of neighbouring groups to the lensing profile in increasingly dense cosmic environments. We show, however, that the observed effect can be entirely attributed to the galaxy density at much smaller scales (within $4 \hm$), which is correlated with the density of the cosmic environments. Within our current uncertainties we find no direct dependence of galaxy halo mass on their cosmic environment.
\end{abstract}




\section{Introduction}

In the standard $\Lambda CDM$ cosmological model, with dark energy and cold dark matter, structure formation in our Universe is described as the gravity-induced growth of small perturbations in the matter density field \cite[]{peebles1970structureform}. This field is dominated by Dark Matter (DM) which outweighs the mass in baryons by a factor  $\sim5$ \cite[]{planck2015}. As a consequence, the properties of baryonic structures are expected to be dominated by the underlying DM density field. More specifically, when over-dense regions undergo gravitational collapse, they form bound structures referred to as DM haloes \cite[]{peebles1974haloes}. Galaxies form in those haloes via the cooling of the gas that falls into the gravitational potential of the DM halo \cite[]{white1978galaxyform}. As a halo grows in mass and size due to smooth accretion and mergers \cite[]{white1991merging}, so does the galaxy that inhabits it (although the detailed properties of galaxies are also affected by baryon-specific processes, such as star formation and feedback from stars and active galactic nuclei). Due to increased clustering of high-mass haloes and the accretion of halo mass through mergers, the DM halo mass is predicted to depend on the presence of other haloes within a few Mpc range \cite[]{bardeen1986, cole1989}. The halo abundance at these small scales is henceforth called the \emph{local density} \cite[]{budavari2003correlation}.

It is possible that the properties of haloes also depend on the density field on scales much larger than the extent of the local structure, known as the large-scale structure (LSS) of the Universe. The universal LSS, as revealed by simulations of large portions of the Universe \cite[e.g.][]{springel2005millenium, schaye2015eagle}, manifests itself as an intricate arrangement of matter density distributions: the sheets of DM that separate large underdense voids intersect to form filaments, which again form dense knots wherever they cross. These structures, collectively called the \emph{cosmic web} \cite[]{bond1986web}, act as a skeleton to large baryonic structures like gas clouds, galaxies, clusters and superclusters. Through the attraction of baryons by DM, large galaxy surveys \cite[e.g.][]{jones20096df, waerbeke2013cfhtlens, tempel2014sdss, garilli2014vipers} are able to observe the cosmic DM web reflected in the large-scale distribution of galaxies.

The question arises whether one can establish a correlation between galaxy halo properties and their location in the cosmic web, independently of the effects of the local environment in which the halo resides. Using numerical simulations, \cite{hahn2009} predicted that the mass of haloes is affected by tidal forces when a large-scale structure resides within 4 virial radii of the halo. According to their simulations these tidal effects can, especially in filaments, suppress halo formation and even extract mass from haloes if they pass the large-scale structure within 1.5 virial radii. On the other hand, they find an increase in the abundance of small haloes near massive structures which, through mergers, can likewise affect halo masses.

The effects of tidal forces on halo formation and mass were also studied by \cite{ludlow2011peaks}. Using $\Lambda CDM$ cosmological simulations they found that, while $\sim70$ percent of the DM haloes should collapse at the location of peaks in the local density field with a mass of similar scale, there should exist a small fraction of haloes that arise from smaller density fluctuations. Compared to regular haloes, these `peakless haloes' should be more strongly affected by tidal forces from neighbouring large-scale structures. However, like \cite{hahn2009}, \cite{ludlow2011peaks} showed that, in the local universe, peakless haloes also reside in denser \emph{local} environments (up to a few Mpc scales).

\cite{eardley2015}, henceforth called E15, classified all galaxies into one of four \emph{cosmic environments}: voids, sheets, filaments and knots. Following \cite{mcnaught2014} they also measured the number density of galaxies within $8 \hm$ radii (\emph{local density}, see Sect. \ref{sec:density}), and found the distribution in local density of galaxies in each cosmic environment. From these local density distributions they concluded that galaxies in denser cosmic environments (e.g. knots) tend to have higher local densities as well. The correlation between large-scale density and the local abundance of haloes complicates observational tests of the predicted tidal effects on galaxy properties. In order to separate the effects of local density from those of the cosmic web, E15 used a `shuffling' method (see Sect. \ref{sec:shuffled}). By creating four new galaxy samples which retain the local density distribution from the original cosmic environments, but with the galaxies shuffled between the cosmic environments, they erased the information from the cosmic environment classification while retaining the information on local density. By comparing the galaxies in these `shuffled environments' to those in the true cosmic environments, they were able to eliminate the dependence on the local overdensity of their measurement of the galaxy luminosity function. In this work we use the environment classification from E15, and follow their shuffling method in order to extract the effect on halo mass from the cosmic environment alone, without effects from the local density. As explained in Sect. \ref{sec:density} we use $4 \hm$ radii to measure the local density, instead of the $8 \hm$ used in E15. This might complicate the comparison of our results with E15, but is necessary due to the different nature of the luminosity function and the halo mass measurement.

The effect of the cosmic web on galaxies was already probed observationally by several groups using different galaxy properties: \cite{alpaslan2015environment} measured the effect of the cosmic web on $u-r$ colour, luminosity, metallicity and morphology of galaxies; \cite{darvish2014cosmic} measured the stellar mass, star formation rate (SFR), SFR-mass relation and specific SFR of galaxies in different cosmic environments; and E15 used their method to measure the galaxy luminosity function. In these and similar studies the importance of the DM haloes of galaxies is often stressed, and the possible effect of the cosmic web on the measured galaxy properties is often expected to be a secondary consequence of the effect on the DM halo. Our goal, therefore, is to perform the first direct measurement of the effect of the cosmic web on galaxy halo mass.

To statistically measure the effect of the cosmic environment on the DM halo mass of galaxies we use weak gravitational lensing, currently the only method that provides a direct measure of the mass of a system without any assumptions on its dynamical state. More specifically, we use galaxy-galaxy lensing \cite[see e.g.][]{brainerd1996ggl, hoekstra2004, mandelbaum2006}: the coherent tangential distortion of background galaxy images due to the gravitational deflection of light by the mass of a sample of foreground galaxies and their haloes. To select foreground galaxies we use the spectroscopic Galaxy And Mass Assembly survey \cite[]{driver2011gama}, whereas the images of the background galaxies are taken from the photometric Kilo-Degree Survey \cite[]{dejong2013kids}. This combination of data sets was also employed by the galaxy-galaxy lensing studies of \cite{viola2015} to measure the masses of galaxy groups, \cite{sifon2015} to study group satellites, and \cite{uitert2016} to measure the stellar-to-halo mass relation. To infer the mass of the haloes associated with the lens galaxies, we employ a simple halo model on the measured galaxy-galaxy lensing signals.

We discuss the lensing methodology and the details of the lens and source samples in Sect. \ref{sec:lensing}. The classification of the cosmic environments and the methods used to correct for the differences in the local density and stellar mass distributions of the galaxy samples are discussed in Sect. \ref{sec:envclass}. In Sect. \ref{sec:analysis} we present the analysis of the lensing profiles in the cosmic environments and the model fitting procedure used to extract the galaxy halo masses from these density profiles. In Sect. \ref{sec:results} we present the resulting masses of DM haloes. Section \ref{sec:discon} contains the discussion and conclusions.

Throughout the paper we adopt the following cosmological parameters: $\Omega_{\rm m}=0.315$, $\Omega_{\rm \Lambda}=0.685$, $\sigma_8 = 0.829$, $n_{\rm s}=0.9603$ and $\Omega_{\rm b} h^2=0.02205$, motivated by \cite{planck2015}. The reduced Hubble constant $h = H_0 / (100 \, {\rm km/s/Mpc})$ is left free or is substituted with 1 where needed.

\section{Galaxy-Galaxy Lensing Analysis}
\label{sec:lensing}

The light from distant galaxies is deflected by density fluctuations along the line of sight. As a consequence, the images of distant galaxies are magnified and distorted (sheared). The technique that measures the weak coherent distortion of a population of background galaxies by a foreground density distribution is called \emph{weak gravitational lensing} (WL), or \emph{galaxy-galaxy lensing} (GGL) when it is used to measure the density distribution around foreground galaxies (lenses) using the shear of many background galaxies (sources) \cite[for an overview, see e.g.][]{bartelmann2001lensing, schneider2006lensing}. These small shape distortions ($\sim1\%$ of the intrinsic galaxy ellipticity) can only be measured statistically by azimuthally averaging the shear of a field of sources around individual lenses, and stacking these lens signals for large samples of foreground galaxies, selected according to their observable properties. The measured quantity is the shear component tangential to the line connecting the lens and source galaxy. Combining this quantity for all lens-source pairs of a lens sample results in the average tangential shear $\langle\gamma_{\rm t}\rangle(R)$, which can be related to the Excess Surface Density (ESD) profile $\Delta \Sigma(R)$. This is defined as the surface mass density $\Sigma(R)$ at the projected radial distance $R$ from the lens centre, subtracted from the average density $\bar \Sigma(<R)$ within that radius:
\begin{equation}
	\langle\gamma_{\rm t}\rangle(R) \Sigma_{\rm crit} = \Delta \Sigma (R) = {\bar \Sigma}(<R) - \Sigma (R) \, .
	\label{eq:deltasigma}
\end{equation}
Here $\Sigma_{\rm crit}$ is the critical density surface mass density:
\begin{equation}
	\Sigma_{\rm crit} = \frac{c^2}{4\pi G} \frac{D(z_{\rm s})}{D(z_{\rm l}) \, D(z_{\rm l}, z_{\rm s})} \, ,
\end{equation}
which is the inverse of the lensing efficiency: a geometrical factor that determines the strength of the lensing effect, depending on the angular diameter distance from the observer to the lens $D(z_{\rm l})$, to the source $D(z_{\rm s})$, and between the lens and the source $D(z_{\rm l}, z_{\rm s})$. In this equation $c$ denotes the speed of light and $G$ the gravitational constant. Our ESD measurement procedure follows Sect. 3.3 of \cite{viola2015}\footnote{One difference between our procedures is the method that decides which $1 \deg^2$ KiDS tiles contribute to the ESD profile of every GAMA foreground galaxy. In \cite{viola2015} the sources within a KiDS tile contributed to the ESD profile of a lens if the projected distance $R_{\rm lt}$ between the lens and the centre of the tile was smaller then the maximal separation $R_{\max}$ to which the ESD profile was measured: $R_{\rm lt} < R_{\max}$. This method was slightly suboptimal, since not all sources contributed to the lensing signal at larger scales. In this paper the method is optimized to encompass the contribution of all KiDS sources to the ESD profile of each lens. We first calculate the projected radius $R_{\rm t}$ of the great circle around each $1 \deg^2$ KiDS tile. The sources within a KiDS tile contribute to the ESD profile of a lens if the radius $R_{\rm t}$ of the circle is smaller then $R_{\rm max}$: $R_{\rm t} < R_{\max}$.}

\subsection{GAMA lens galaxies}
\label{sec:gama}
The positions of the foreground lenses used for our GGL study are selected from the Sloan Digital Sky Survey \cite[SDSS]{abazajian2009sdss}, and have redshifts measured by the Galaxy And Mass Assembly \cite[hereafter GAMA,][]{driver2011gama} survey, a spectroscopic survey on the Anglo-Australian Telescope with the AAOmega spectrograph. We use the GAMA galaxy catalogue resulting from the three equatorial regions (G09, G12 and G15) of the final GAMA data release \cite[GAMA II,][]{liske2015gamaII}. These equatorial regions span a total of $\sim180 \deg^2$ and are $98\%$ complete down to a Petrosian $r$-band magnitude of $m_{\rm r} = 19.8$. This catalogue contains $180,960$ galaxies, of which we use the sample of $\sim113,000$ galaxies within the redshift range $0.039 < z_{\rm l} < 0.263$ (see Sect. \ref{sec:density}) with a high-quality redshift measurement ($nQ \geq 3$) as lenses. Since $\sim55\%$ of the GAMA area is covered by the Kilo-Degree Survey data that we use for this analysis, $\sim80 \%$ of these galaxies have at least some overlap with the available area (see Sect. \ref{sec:kids}), and therefore contribute (in varying degrees) to the lensing signal. This amounts to a total of $91195$ galaxies contributing to the lensing signal.

In Sect. \ref{sec:mstarweight} of this paper we make use of the stellar masses of the GAMA galaxies, which are measured by \cite{taylor2011mstar} by fitting \cite{bruzual2003mstar} stellar population synthesis models to the \emph{ugriz} observations of the SDSS. The stellar masses are corrected for flux falling outside the automatically selected aperture using the `flux-scale' parameter \cite[following the procedure discussed in ][]{taylor2011mstar}. Galaxies without a well-defined stellar mass or fluxscale correction are removed from our sample.

In Sect. \ref{sec:halomodel} we use the classification of GAMA galaxies into galaxy groups, in order to accurately model the contribution of different galaxies to the GGL signal. We use the classification of galaxies into groups as listed in the 7th GAMA Galaxy Group Catalogue by \cite{robotham2011lenscat}. The galaxies in the GAMA II catalogue are classified as either the central or a satellite of their group, using the Friends-of-Friends (FoF) group finding algorithm described in \cite{robotham2011lenscat}. The FoF algorithm groups galaxies depending on the distances to each other in both projected and line-of-sight space. The projected and line-of-sight linking lengths are calibrated against mock catalogues. These mocks are also used to test that global properties of groups, such as their total number, are well recovered by the algorithm. The FoF method also finds galaxies that do not belong to any group (non-group galaxies). Note that some non-group galaxies might actually be centrals of groups with satellites that fall below the visible magnitude limit. Conversely, non-group galaxies can erroneously be classified as group members due to projection effects. However, this misidentification is primarily expected to happen at high redshifts, whereas our sample only contains galaxies up to redshift $z_{\rm l} = 0.263$. Also note that some galaxies classified as satellites might actually be centrals, and some satellites might be erroneously identified as non-group galaxies. This misidentification is most common for the smallest groups (with less than 5 members). Since we primarily use the group classification to account for the contribution of nearby galaxies to the GGL signal, it is of little consequence whether these galaxies are classified as satellites or neighbouring group centrals since both are accounted for in the model. Furthermore, the GGL analysis of \cite{uitert2016} to determine the fraction of satellites in the central galaxy sample, shows that the satellite fraction of GAMA is accurate for galaxies with a stellar mass up to $\sim10^{11} \msun$. For these reasons, it safe to use galaxies with a small number of members in our analysis. The definition of the central galaxy used in this paper is the Brightest Central Galaxy (BCG). In \cite{viola2015} the BCG is shown to be the most accurate tracer of the centre of a group halo (together with the iteratively selected central galaxy).

\subsection{KiDS source galaxies}
\label{sec:kids}

The background sources used to measure the GGL profiles around the lenses are observed with the Kilo-Degree Survey \cite[hereafter KiDS,][]{dejong2013kids}, a \emph{ugri} photometric survey on the VLT Survey Telescope \cite[]{capaccioli2011vlt} using the OmegaCAM wide-field imager \cite[]{kuijken2011omegacam}. We use the $109 \deg^2$ area of the publicly available KiDS-DR1/2 data release \cite[]{dejong2015dr12, kuijken2015lensingkids} that overlaps with the equatorial GAMA areas. With the masks of bright stars and image defects applied, this amounts to a total of 68.5 $\deg^2$ of KiDS area that overlaps with the GAMA survey.

For the measurements of the source ellipticities we use the $r$-band data, which have a median seeing of $0.7\arcsec$ and a mean point spread function (PSF) ellipticity of 0.055. The $r$-band images are first reduced with the {\scshape Theli} pipeline \cite[]{erben2013theli}. The sources are then detected from the stacked images by {\scshape SExtractor} \cite[]{bertin1996sextractor}. For each detected source, the Bayesian \emph{lens}fit method \cite[]{miller2013lensfit} measures the ellipticities $\epsilon_1$ and $\epsilon_2$ with respect to the equatorial coordinate frame, together with a weight $w_{\rm s}$ which is related to the uncertainty on the ellipticity measurements. The corresponding effective number density of sources is $n_{\rm eff} \approx \frac{\sigma_{\rm SN}^2}{A}\sum_{s} w_s = 4.48 \rm gal/arcmin^2$, where $A$ is the area and $\sigma_{\rm SN} = 0.255$ the intrinsic ellipticity dispersion (shape noise) \cite[]{kuijken2015lensingkids}.

The photometric redshifts of the sources are derived from all bands, which are first processed and calibrated using the Astro-WISE optical image pipeline \cite[]{mcfarland2013astrowise}. The Gaussian Aperture and PSF \cite[GAaP,][]{kuijken2008gaap} method measures the matched aperture colours of the sources, corrected for differences in seeing. These are in turn used to determine the photometric redshifts through the Bayesian Photometric Redshift pipeline (BPZ, \citealp{benitez2000bpz} following \citealp{hildebrandt2012bpz}). The source redshift probability distribution $p(z)$ is sampled using 70 linearly spaced source redshift bins between $0 < z_{\rm s} < 3.5$. We use the full photometric redshift probability distribution $p(z_{\rm s})$ returned by the BPZ to calculate the critical surface density for each lens-source pair:
\begin{equation}
	\label{eq:sigmacrit}
	\Sigma_{\rm crit}^{-1} = \frac{4\pi G}{c^2} D(z_{\rm l}) \int_{z_{\rm l}}^{\infty} \frac{D(z_{\rm l}, z_{\rm s})}{D(z_{\rm s})} p(z_{\rm s}) \: {\rm d} z_{\rm s} \, ,
\end{equation}
where the integral is over the part of the $p(z_{\rm s})$ where the source redshift bins $z_{\rm s}$ are greater than the redshift $z_{\rm l}$ of the lens. Only sources with a $p(z)$ peak within $0.005 \leq z_{\rm B} < 1.2$, where the summed $p(z)$ posteriors agree well with the spectroscopic redshift distribution \cite[]{kuijken2015lensingkids}, are used for the analysis.

We assign a weight $W_{\rm ls}$ to each lens-source pair, that incorporates the ellipticity measurement weight $w_{\rm s}$ (from \emph{lens}fit) which down-weights lens-source pairs that have less reliable shape measurements, as well as the lensing efficiency $\Sigma_{\rm crit}^{-1}$ which down-weights lens-source pairs that are very close or distant in redshift, and are therefore less sensitive to lensing. The total weight is defined as:
\begin{equation}
	W_{\rm{ls}} = w_{\rm s} \Sigma_{\rm crit}^{-2} \, .
	\label{eq:weights}
\end{equation}

We apply an average multiplicative calibration $1+K(R)$ to the stacked ESD profile, in order to account for the noise bias in the shape measurement \cite[]{melchior2012noisebias, heymans2012cfhtlens}. The bias contribution $m_{\rm s}$ of each source is estimated from a \emph{lens}fit analysis of simulated images \cite[]{miller2013lensfit}. For every radial bin $R$ we determine the average correction:
\begin{equation}
	K(R) = \frac{\sum_{ls} W_{ls} m_{s} }{ \sum_{ls} W_{ls} } \, ,
	\label{eq:biascorr}
\end{equation}
which has a value of $K(R)\sim0.1$ for all considered values of $R$. In addition to an average multiplicative calibration, we apply an additive calibration term $c_{\rm s}$ to each separate ellipticity value. See \cite{kuijken2015lensingkids} for more information on these calibrations.

The ESD profile $\Delta \Sigma(R)$ from Eq. (\ref{eq:deltasigma}) can be measured by computing the tangential ellipticity $\epsilon_{\rm t}$:
\begin{equation}
	\epsilon_{\rm t} = -\epsilon_1 \cos(2\phi) - \epsilon_2 \sin(2\phi) \, ,
\end{equation}
where $\phi$ is the angle between the source and the lens centre. The tangential ellipticity is measured for every lens-source pair in a range of circular bins at radius $R$ with width $\delta R$, and the resulting signal is divided by the multiplicative calibration term to arrive at the ESD profile:
\begin{equation}
	\Delta \Sigma (R) = \frac{1}{1+K(R)} \frac{\sum_{ls} W_{ls} \epsilon_{\rm t} \Sigma_{\rm crit} }{ \sum_{ls}{W_{ls}} } \, .
	\label{eq:ESDmeasured}
\end{equation}

The uncertainty on the measured ESD profile corresponds to the square root of the diagonal of its analytical covariance matrix. As detailed in Sect. 3.4 of \cite{viola2015}, we compute the analytical covariance of the contributions to the ESD signal from each separate source that contributes to the stacked ESD profile of the lens sample. This covariance is not only computed between each radial bin, but also between the different stacked lens samples. These off-diagonal covariance elements are used within the model fitting procedure (see Sect. \ref{sec:halomodel}). Section 3.4 of \cite{viola2015} shows that the error bars from the analytical covariance are in agreement with the bootstrap error bars from $\sim100$ KiDS tiles, up to projected radius $R=2\hm$. The lensing signal around random points is not consistent with zero beyond this projected radius, due to the patchiness of the survey area. Below $20\hk$ the decreasing number of sources and increasing contamination from foreground galaxy light also result in unreliable measurements. We therefore compute the ESD profile for 10 logarithmically spaced radial bins within $0.2 < R < 2 \hm$.

\section{Environment Classification}
\label{sec:envclass}

\subsection{Cosmic environments}
\label{sec:environment}

The goal of this work is to study the mass of galaxy haloes as a function of their location in the cosmic web, henceforth called their \emph{cosmic environment}. In E15 the entire volume of the GAMA survey is classified into four different cosmic environments: voids, sheets, filaments and knots. These environments are identified following the tidal tensor prescription of \cite{hahn2007environment}, which classifies each Cartesian $R_{\rm c} = 3 \hm$ volume element (cell) in the GAMA survey into one of these four cosmic environments, based on the number of gravitationally collapsing dimensions of that cell. A volume element can be collapsing in 0, 1, 2 or 3 dimensions, and is respectively classified as belonging to a void, sheet, filament or knot.

To determine the number of collapsing dimensions of each cell, E15 use the number density of galaxies in the $R_{\rm c} = 3 \hm$ Cartesian grid. This discrete density field is smoothed by applying a Gaussian filter with a width $\sigma_{\rm s}$, resulting in the total effective smoothing scale $\sigma^2 = R^2_{\rm c}/6 + \sigma^2_{\rm s}$. From this smoothed density field E15 derive the gravitational potential, which is used to calculate the tidal tensor of each cell. Since the tidal tensor is a criterion for a cell's gravitational stability, a dimension of a cubic cell is considered to be collapsing if the corresponding real eigenvalue of the tidal tensor lies below a threshold value $\lambda_{\rm th}$. Depending on its number of collapsing dimensions, each cell is allocated a cosmic environment. Correspondingly E15 assign each galaxy in the GAMA catalogue to the environment of the cell it inhabits, allowing us to study these galaxies and their DM haloes as a function of their cosmic environment.

\begin{figure*}
	\includegraphics[width=\textwidth]{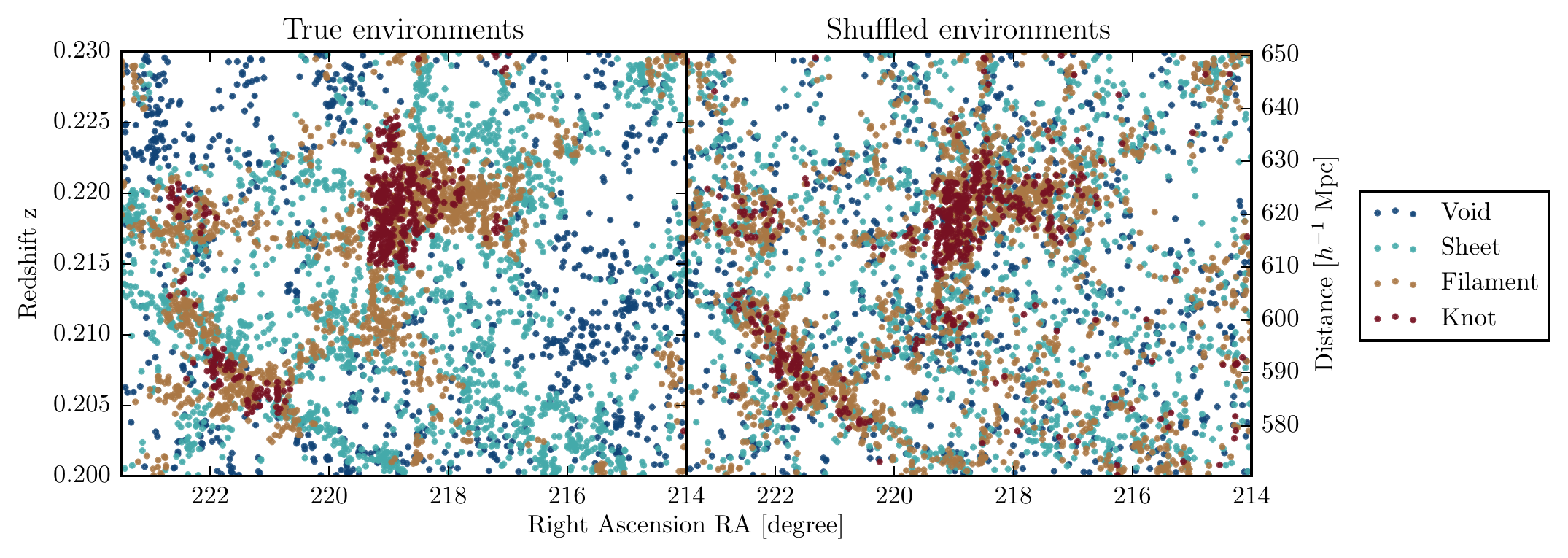}
	\caption{The spatial distribution of galaxies, in Right Ascension (x-axis) and redshift (left y-axis) or comoving distance (right y-axis), distinctly shows the cosmic web structure. This figure presents a slice of $5^\circ$ in Declination (DEC) of a small representative part of the GAMA volume. In the left panel, the colours indicate the classification of galaxies into four cosmic environments: voids (blue), sheets (cyan), filaments (yellow), and knots (red), based on the number of spatial dimensions in which their region is collapsing (E15). In the right panel, the colours represent the corresponding shuffled environments, which only share the local density distribution of the true environments.}
	\label{fig:environments}
\end{figure*}

The values of the effective smoothing scale $\sigma_{\rm s}$ and the eigenvalue threshold $\lambda_{\rm th}$ can be chosen freely in this method, and together determine the number of galaxies that are assigned to each cosmic environment. In order to divide the number of GAMA galaxies as equally as possible among the four cosmic environments, E15 chose $\lambda_{\rm th}=0.4$ and $\sigma_{\rm s}$ such that $\sigma=4\hm$, because these values minimized the root-mean-square dispersion between the fraction of galaxies assigned to each of the cosmic environments. This equal division of galaxies was necessary to ensure a sufficiently high Signal-to-Noise (SN) ratio of measurements in each of the four environments. Although we recognize they are not physically motivated, we adopt the same values of $\lambda_{\rm th}$ and $\sigma$ as E15 for comparison purposes. Furthermore, our analysis likewise benefits from sufficient signal in each cosmic environment, although in our case this does not only depend on the number of lenses but also on the mass of the galaxy haloes. The total number of galaxies in each cosmic environment can be found in Table \ref{tab:lenses}. The left panel of Fig. \ref{fig:environments} gives a visual impression of the spatial distribution of galaxies in the different cosmic environments.

\begin{table}
	\centering
	\caption{The number, mean redshift, mean stellar mass and satellite fraction of the galaxies in each cosmic environment. Note that only $\sim80\%$ of these galaxies overlap with the KiDS area, and therefore contribute to the GGL signal. The values of $\langle M_* \rangle$ are displayed in units of $[10^{10} \msun]$.}
	\begin{tabular}{lllll}
		\hline
	                      & $N$ & $\langle z \rangle$ & $\langle M_* \rangle$ & $f_{\rm sat}$      \\ \hline
		Void              & 19742       & 0.161			      & 2.767       	      & 0.146              \\ 
		Sheet             & 37932       & 0.169               & 3.465         		  & 0.243              \\ 
		Filament          & 41753       & 0.165               & 3.945        		  & 0.363              \\ 
		Knot              & 13457       & 0.157               & 4.354        		  & 0.502              \\ \hline
		Shuffled void     & 19742       & 0.160               & 2.590        		  & 0.174              \\ 
		Shuffled sheet    & 37932       & 0.165               & 3.393        		  & 0.250              \\ 
		Shuffled filament & 41753       & 0.167               & 4.048        		  & 0.350              \\ 
		Shuffled knot     & 13457       & 0.165               & 4.499        		  & 0.484              \\ \hline
		\end{tabular}
	\label{tab:lenses}
\end{table}

\subsubsection{Stellar mass weights}
\label{sec:mstarweight}

\begin{figure}
	\includegraphics[width=\columnwidth]{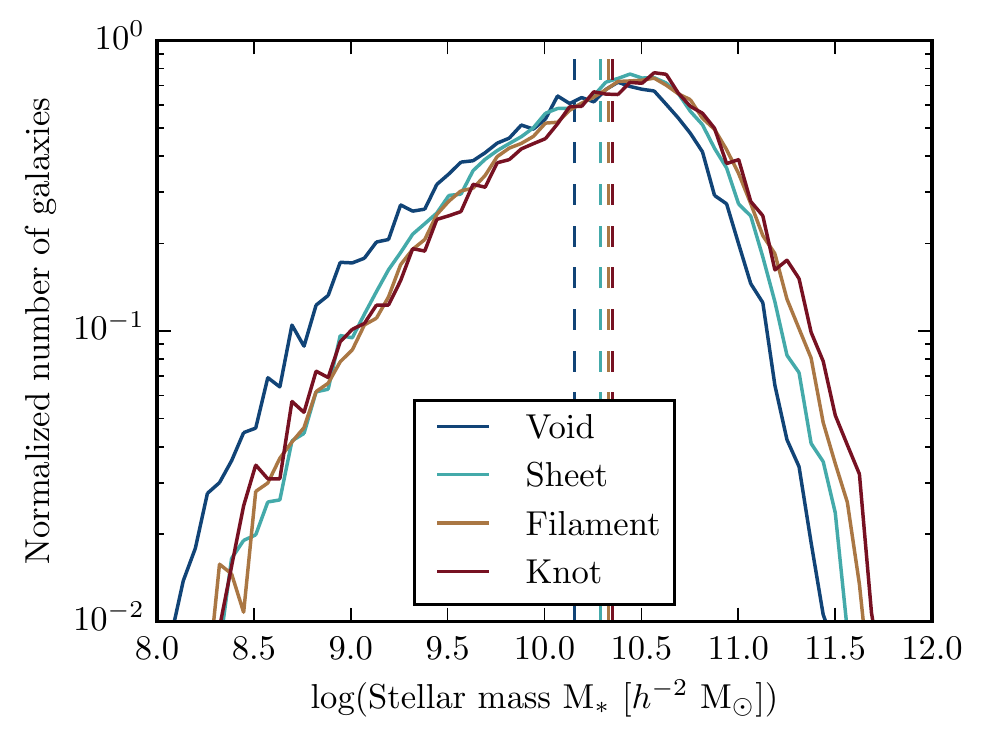}
	\caption{The normalized number of galaxies in each cosmic environment as a function of stellar mass $M_*$. Each vertical dashed lines shows the mean of the $M_*$ distribution. The distributions show that galaxies in denser cosmic environments tend to have slightly higher stellar masses.}
	\label{fig:mstarhist}
\end{figure}

For each of the four cosmic environments, the normalized stellar mass ($M_*$) distribution of galaxies is slightly different. As shown in Fig. \ref{fig:mstarhist}, galaxies in denser environments tend to have higher stellar masses, and voids tend to have lower-mass galaxies ($\log_{10}(\frac{M_*}{\msun}) < 9.5$) compared to the other cosmic environments. Because there exists a correlation between $M_*$ and halo mass \cite[e.g.][]{mandelbaum2006,moster2010,uitert2016}, this difference should be corrected for in order to find the unbiased dependence of halo mass on cosmic environment. To this end we assign a stellar mass weight $w_*$ to each lens, which is used to weigh the contribution of that lens to the stacked GGL profile. For 100 linearly spaced bins in $\log_{10}(M_*)$, we count the number of lenses $N(M_*, E)$ in each cosmic environment $E$. This is compared to the average number of galaxies $\langle N\rangle(M_*)$ in all environments that reside in the corresponding $M_*$ bin, in order to find the stellar mass weight:
\begin{equation}
w_*(M_*, E) = \frac{\langle N\rangle(M_*)}{N(M_*, E)} \, ,
\end{equation}
which is assigned to all galaxies in that $M_*$ bin and environment. The stellar mass weight $w_*$ is applied to each galaxy's contribution to the ESD profile through Eq. (\ref{eq:ESDmeasured}), such that it becomes:
\begin{equation}
\Delta \Sigma_* (R) = \frac{1}{1+K_*(R)} \frac{\sum_{l} w_* \sum_{s} W_{ls} \epsilon_{\rm t} \Sigma_{\rm crit} }{\sum_{l} w_* \sum_{s} {W_{ls}} } \, .
\end{equation}
where the average multiplicative bias correction from Eq. (\ref{eq:biascorr}) has become:
\begin{equation}
K_*(R) = \frac{\sum_{l} w_* \sum_{s}  W_{ls} m_{s} }{\sum_{l} w_* \sum_{s}  W_{ls} } \, ,
\end{equation}
Likewise the lens weight is incorporated into the uncertainty through the calculation of the analytical covariance matrix (see Sect. \ref{sec:kids}). In this way, we give higher weights to galaxies with a stellar mass that is under-represented in a specific environment compared to the average of the four environments. However, because the $M_*$-distributions are similar in our case, there is only a small difference between the stacked ESD profiles with or without the stellar mass weights, and we can use this correction as a reasonable approximation.

\subsection{Local density}
\label{sec:density}

A complicating factor in studying the dependence of halo mass on cosmic environment, is that denser cosmic environments also have a higher average density at smaller (a few Mpc) scales: the \emph{local density}. In order to determine the effects of the cosmic environment independent of local influences, we need to define and measure the local densities of galaxies and correct for them. We measure the local density $\rho_{\rm r}$ following \cite{mcnaught2014}, who define this quantity from the number of tracer galaxies $N$ inside a sphere of co-moving radius $r$ around a galaxy. Following \cite{croton2005density} the tracers used for measuring $N$ belong to a `volume limited Density Defining Population' (DDP): the sample of galaxies that is visible over the entire range in redshift $z_{\rm l}$, given a certain cut in $r$-band absolute magnitude $M^{\rm h}_{\rm r} = M_{\rm r} - 5 \log_{10}(h)$ (with the k-correction and luminosity evolution correction applied). \cite{mcnaught2014} apply a narrow cut in absolute magnitude: $-21.8 < M^{\rm h}_{\rm r} < -20.1$, in order to preserve a relatively wide redshift range: $0.039 < z_{\rm l} < 0.263$. Following this $M^{\rm h}_{\rm r}$ and $z_{\rm l}$ cut, we obtain a DDP containing 44317 GAMA galaxies.  We count the number of tracers $N_{\rm DDP}$ in a sphere around GAMA galaxies to determine their local density:
\begin{equation}
\rho_{\rm r} = \frac{N_{\rm DDP}}{\frac{4}{3} \pi r^3} \frac{1}{C_{\rm v} C_{\rm z}} \, ,
\end{equation}
where $C_{\rm v}$ is the volume correction accounting for the fraction of the sphere lying outside the boundaries of the survey or redshift cut, and $C_{\rm z}$ accounts for the redshift completeness of the volume (measured using the GAMA masks).

In order to determine the local overdensity $\delta_{\rm r}$ within co-moving radius $r$ around a galaxy, we compare $\rho_{\rm r}$ to the mean DDP number density $\bar{\rho}$ over the full GAMA volume:
\begin{equation}
\delta_{\rm r} = \frac{\rho_{\rm r}-\bar{\rho}}{\bar{\rho}} \, .
\end{equation}
When corrected for redshift completeness using the GAMA masks, the total volume (within the designated redshift range) of the three equatorial GAMA fields is $V_{\rm GAMA} = 7 \times 10^6 (\hm)^3$, resulting in an effective mean DDP galaxy density of $\bar{\rho} = 6 \times 10^{-3} (\hm)^{-3}$.

\begin{figure}
	\includegraphics[width=\columnwidth]{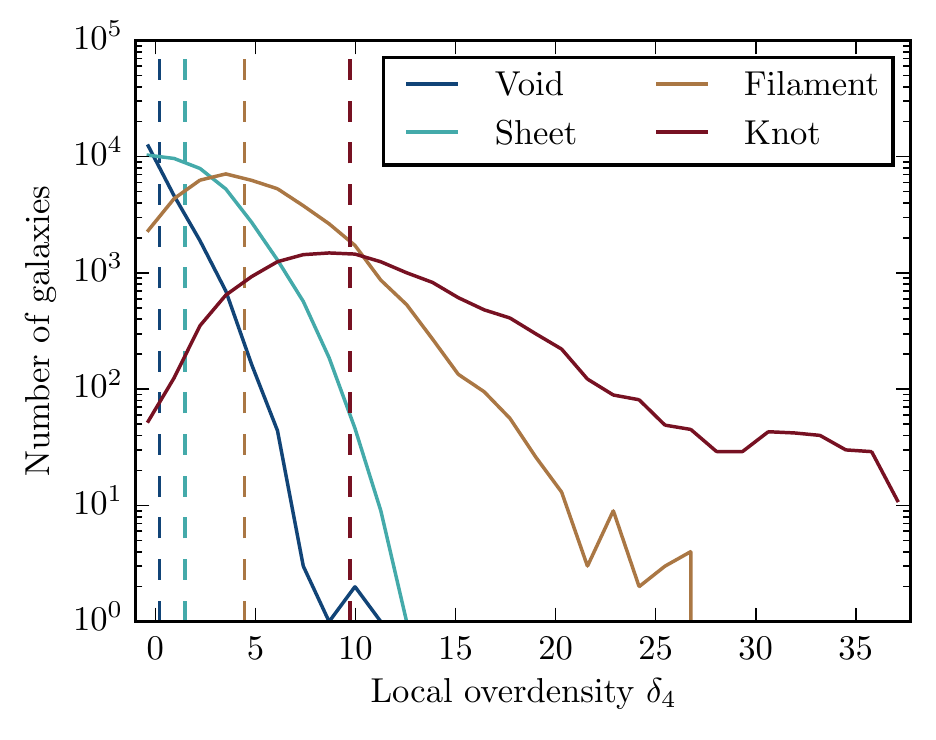}
	\caption{The number of galaxies in each environment as a function of local overdensity $\delta_4$ (overdensity within a $4\hm$ radius). Each vertical dashed line shows the mean of the $\delta_4$ distribution. As expected galaxies in denser cosmic environments tend to have higher local overdensities, although there remains significant overlap between the distributions.}
	\label{fig:envhist}
\end{figure}

Using the DDP we measure the value of $\delta_4$, the overdensity within $r = 4 \hm$, for \emph{all} GAMA galaxies within the redshift range of the DDP (including those outside the absolute magnitude range), amounting to a sample of $\sim113,000$ lenses. We choose spheres with $r = 4 \hm$ to probe local overdensities at the scale of the correlation length of the LSS \cite[]{budavari2003correlation}, which is also the smallest possible scale that still avoids major problems related to scarce tracer galaxies and redshift space distortion on small scales \cite[]{croton2005density}. For each cosmic environment we find a different distribution in $\delta_4$, as shown in Fig. \ref{fig:envhist}. Not surprisingly denser environments (e.g. knots) contain more galaxies with high local overdensity, while sparser environments (e.g. voids) have lower overdensities. Note, however, that there exists a significant overlap between the different overdensity distributions. This overlap allows us to separate the effect of the cosmic environment on the ESD profile from the effect of local overdensity, enabling us to study the direct dependence of the cosmic environment on halo mass. By shuffling galaxies between the cosmic environments while keeping the local overdensity distribution the same, we create so-called shuffled environments.

\subsection{Shuffled environments}
\label{sec:shuffled}

To account for the different local density distributions in each cosmic environment, we follow E15 in creating a set of four \emph{shuffled environments}: galaxy samples that retain the local overdensity distribution of the true cosmic environments, but contain galaxies that are randomly selected from all cosmic environments, effectively erasing the information from the environment classification. By comparing the galaxies in each shuffled environment to those from the corresponding true environment, we are able to eliminate any dependence on the local overdensity, and extract the effects of the cosmic environment alone.

\begin{figure}
	\includegraphics[width=\columnwidth]{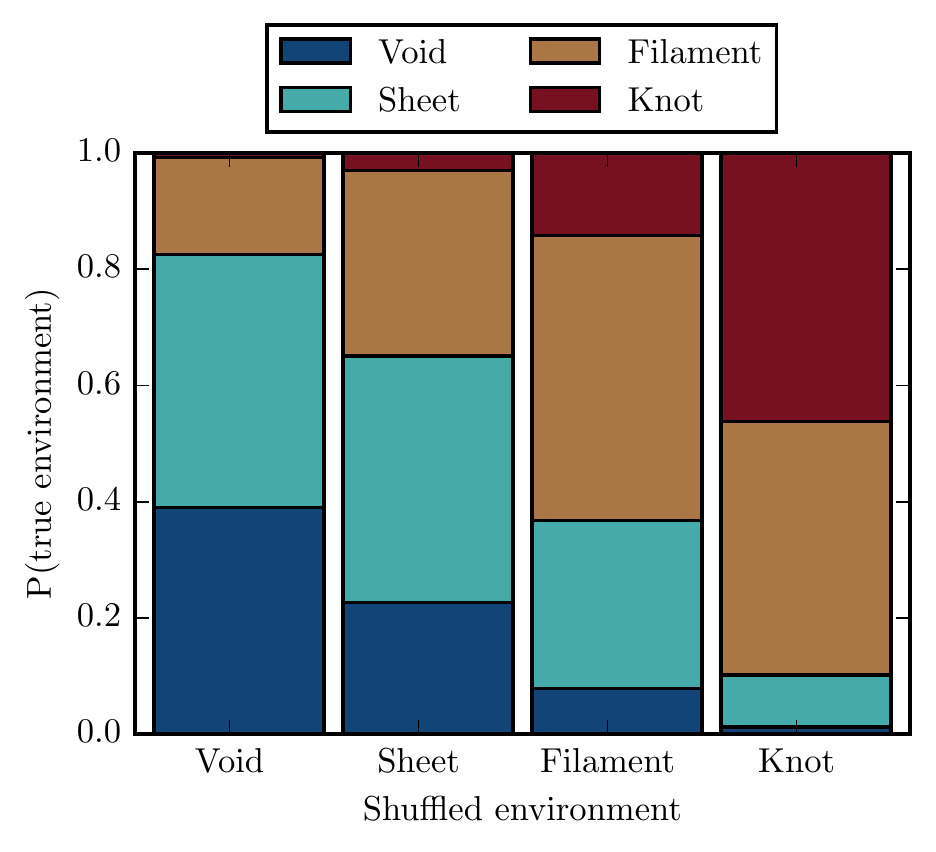}
	\caption{The proportion of galaxies from each true cosmic environment that is present in each shuffled environment. As expected a large fraction of galaxies in each shuffled environment originates from the same true environment, although that fraction is $< 0.5$ for all shuffled environments.}
	\label{fig:envprop}
\end{figure}

In practice, all galaxies are divided into 100 $\delta_4$ bins. For each true cosmic environment we create a shuffled environment, by assigning the same number of galaxies in each $\delta_4$ bin to the corresponding shuffled environment. These galaxies, however, are randomly selected from the full sample, and could therefore be residing in any cosmic environment. Randomly selected galaxies from a high $\delta_4$ bin will be more likely to reside in knots than in voids (due to the correlation between local density and cosmic environment), but every shuffled environment contains a distribution of galaxies from different true environments due to the overlapping $\delta_4$ distributions (see Fig. \ref{fig:envhist}). The proportion of galaxies from true cosmic environments residing in each shuffled environment can be seen in Fig. \ref{fig:envprop}, which shows that up to half of the galaxies in each shuffled environment originate from the same true cosmic environment. In the right panel of Fig. \ref{fig:environments} we show the spatial distribution of galaxies in different shuffled environments, which is likewise correlated with the distribution of galaxies in true cosmic environments shown in the left panel. Although the correlation between the true and shuffled environments complicates the detection of a direct effect from cosmic environments, the relationship between cosmic environment and local density cannot be circumvented in another way without significantly reducing the lens sample. Furthermore, selecting a very different galaxy sample or shuffling method complicates the comparison with the results from E15. We can slightly reduce the proportion of knot galaxies in knots by removing all galaxies with $\delta_4 > 15$ from our sample, but this small effect does not significantly affect our results.

\section{Analysis of the lensing profiles}
\label{sec:analysis}

\subsection{Contributions of group samples}
\label{sec:groupmembers}

\begin{figure}
	\includegraphics[width=\columnwidth, angle=0]{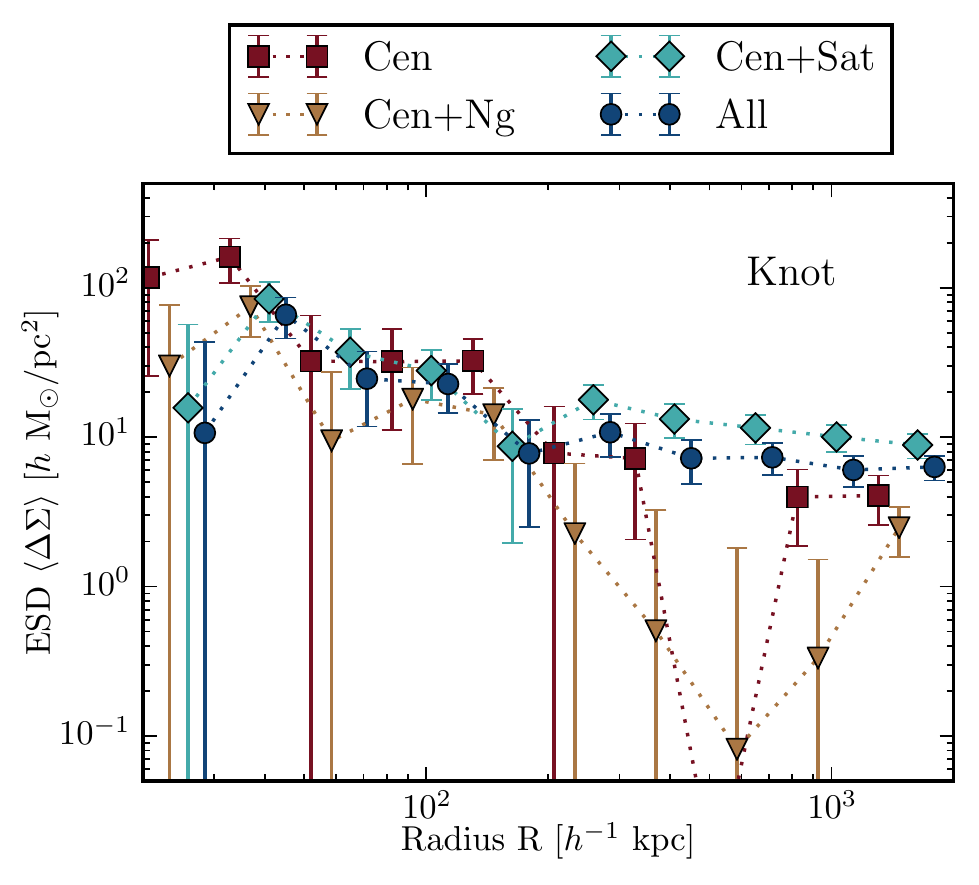}
	\caption{The ESD profiles (with $1\sigma$ error bars) of GAMA galaxies in the knot environment, stacked according to their group membership and weighted to correct for differences in the stellar mass distribution. The different ESD profiles correspond to four group samples: centrals only (Cen), centrals and non-group (Cen+Ng), centrals and satellites (Cen+Sat), and all galaxies (All: centrals, satellites and non-group). The dotted lines are used to guide the eye between data points of the same group sample.}
	\label{fig:ESD_envS4_ranks}
\end{figure}

To obtain the ESD profile of galaxies in each cosmic environment, we stack the lensing signals of these galaxies as detailed in Sect. \ref{sec:kids}. The interpretation of this stacked ESD profile is complicated by the fact that galaxies, apart from residing in a cosmic environment, may also belong to a galaxy group. We use the 7th GAMA Galaxy Group Catalogue \cite[$\rm G^3Cv7$,][]{robotham2011lenscat} to identify the group classification of the lenses that contribute to the stacked ESD profile of each cosmic environment. In Fig. \ref{fig:ESD_envS4_ranks} we show the contribution of different galaxy selections to the total ESD profile of galaxies in the knot environment, where we find the contribution from satellites and neighbouring galaxies to be the largest. We show the signal for central galaxies only, and add the contribution from non-group galaxies, satellite galaxies, or both (all galaxies). The correction for the difference in the stellar mass distribution of the cosmic environments has been applied during the stacking procedure (see Sect. \ref{sec:mstarweight}). The GGL profile in knots shows that, after the first radial bin, the ESD is consistent for all lens samples at scales $R < 200\hk$, where the haloes of the stacked galaxies themselves dominate (as opposed to haloes of neighbouring galaxies). Within the first bin we see a hint of the expected difference between central, non-group and satellite galaxy masses (in order of expected mass), although the differences stay within $1 \sigma$. However, at $R > 200\hk$ the GGL signal changes significantly with the addition of satellite galaxies to the stack. Where the ESD profiles of lens samples without satellites drop sharply, the profiles of samples with satellites does not, due to the off-set contribution of the satellites' host haloes \cite[also seen in][]{sifon2015}. These changes in the ESD profile imply that, as the contributions from different group members are added to the stacked signal, we need to model these different components to account for the total lensing signal. This complicates the interpretation of differences between the ESD profiles in the cosmic environments, including halo mass estimates.

\begin{figure}
	\includegraphics[width=\columnwidth]{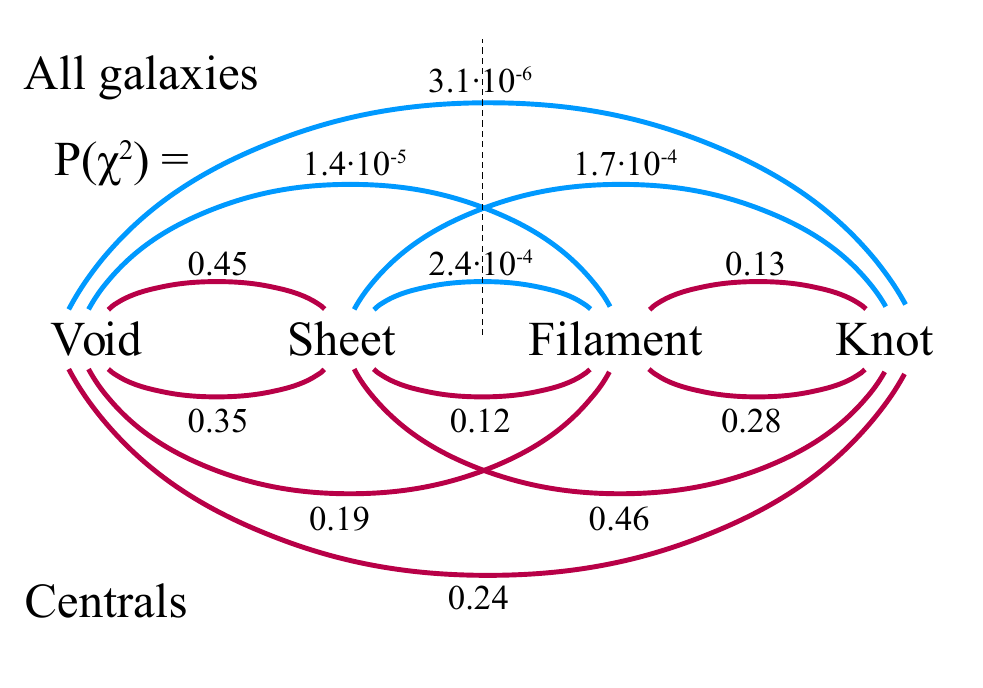}
	\caption{The resulting $P(\chi^2)$ values of the $\chi^2$ independence test between the ESD profiles $\Delta\Sigma(R)$ from different cosmic environments, shown for two galaxy samples: all galaxies (top) and centrals only (bottom). When the probability $P(\chi^2)$ to draw the $\Delta\Sigma(R)$ values in question from the same normal distribution is less than $0.05$, the line between those environments is coloured blue (independent), otherwise red (dependent). The dotted vertical line emphasises the fact that the difference between two environments is significant when it crosses the border between sheet and filament. This figure shows that the stacked ESD profiles of only centrals do not show any significant difference between cosmic environments, whereas the stacked profiles of all galaxies do.}
	\label{fig:diff_diagram}
\end{figure}

Although the sample containing only group centrals is the simplest to model, the low SN ratio of the lensing signal might prohibit the analysis if the SN ratio is too low to even find the expected difference between the four cosmic environments (as measured in e.g. E15), let alone a difference between true and shuffled environments. In order to find whether this is the case, we apply a $\chi^2$ independence test to the ESD profiles $\Delta\Sigma$ in different environments E1 and E2:
\begin{equation}
\chi^2 = \sum_{R}\frac{(\Delta\Sigma^{\rm E1}_R - \Delta\Sigma^{\rm E2}_R)^2}{(\sigma^{\rm E1}_R)^2 + (\sigma^{\rm E2}_R)^2} \, ,
\label{eq:chi2}
\end{equation}
where the index $R$ sums over the radial bins, and $\sigma$ is the uncertainty on $\Delta\Sigma$ calculated from the analytical covariance matrix. We calculate the probability $P(\chi^2)$ to draw the $\Delta\Sigma$ values in question from the same normal distribution, by evaluating the cumulative normal distribution function with $10-1=9$ degrees of freedom (based on the 10 radial bins) at $\chi^2$. We consider the difference between the ESD profiles in two cosmic environments to be significant if $P(\chi^2) < 0.05$. Figure \ref{fig:diff_diagram} shows a diagram of the result of the $\chi^2$ independence test for two samples: `all galaxies' (top) and only `centrals' (bottom). For the sample of centrals there is no measurable difference between the ESD profiles from any of the environments. For the sample of all galaxies there is also no significant difference between voids and sheets, or between filaments and knots; two combinations that can be considered to be `adjacent' in density space. However, there is a measurable difference between sheets and filaments which are likewise adjacent. In fact, every comparison that `crosses' the dotted vertical line between sheets and filaments results in a measurable difference.

In conclusion, this test shows that the difference between the four environments cannot be detected when the stacked ESD profiles contain only the contributions from central galaxies. As a result, we need to add and model the contribution from multiple group members: centrals, satellites and non-group galaxies, in order to measure the average halo mass of galaxies in different cosmic environments.

In this test we can include all or a subset of our 10 radial bins between $20 \hk < R < 2000 \hk$. We therefore repeat this analysis to test whether the difference between the ESD profiles in the cosmic environments is more significant at small scales (by summing over the five innermost radial bins) or large scales (by summing over the five outermost radial bins). Through this test we found that, when all galaxies contribute to the signal, the difference between the ESD profiles in the four cosmic environments is primarily driven by the large scales, indicating that satellites and neighbouring haloes have a major effect on the ESD profiles.

\subsection{Surface density model}
\label{sec:halomodel}

In order to extract the average halo masses from the ESD profiles, we construct a mathematical description of the main contributions to the ESD profile in terms of the stellar and DM components of different galaxy group members, based on the classification by \cite{robotham2011lenscat} of GAMA galaxies into central, satellite and non-group galaxies. Although they have no (visible) satellites we treat non-group galaxies as centrals of groups, which means the two main components of the model correspond to the mass contribution from `centrals' (real centrals and non-group galaxies) and satellites, both residing in the main host halo (which corresponds to the halo of the `central'). Following the prescription of \cite{wright2000model}, we model the DM contribution of both centrals and satellites by an NFW profile \cite[]{navarro1995nfw}:
\begin{equation}
\rho_{\rm NFW}(r) = \frac{\delta_{\rm c} \rho_{\rm m}(\langle z \rangle)}{(r/r_{\rm s})(1+r/r_{\rm s})^2}  \, ,
\label{eq:nfw}
\end{equation}
where $r_{\rm s}$ is the scale radius and $\rho_{\rm m}(\langle z \rangle)$ is the mean density of the Universe, which depends on the mean redshift $\langle z \rangle$ of the lens sample as:
\begin{equation}
\rho_{\rm m} = 3H_0^2(1+ \langle z \rangle)^3\Omega_{\rm m}/(8\pi G) \, .
\end{equation}
The dimensionless amplitude is related to the concentration $c=r_{200}/r_{\rm s}$ via:
\begin{equation}
\delta_{\rm c} = \frac{200}{3} \frac{c^3}{\ln(1+c)-c/(1+c)} \, .
\end{equation}
We include two free parameters in the model: the concentration normalization $f_{\rm c}$, which is the normalization of the \cite{duffy2008massrelation} mass-concentration relation\footnote{We realize that this mass-concentration relation is slightly dated compared to that of e.g. \cite{dutton2014massrelation}, which is based on the \cite{planck2014} cosmology. However, we follow the earlier KiDS-GAMA lensing papers \cite[]{viola2015,sifon2015,uitert2016}, to which our results could then be compared. Furthermore, because of the weak relation of $c(M_{200}, z)$ on $M_{200}$, and the relatively small $z$-range of the GAMA galaxies, the final expression for the concentration will be dominated by the normalization parameter $f_{\rm c}$, and not by the chosen mass-concentration relation.}
\begin{equation}
c(M_{200}, z) = 10.14 f_{\rm c} \left(\frac{M_{200}}{2\times10^{12} \msun}\right)^{-0.089} (1+z)^{-1.01} \, ,
\end{equation}
and the halo mass $M_{200}$ which is defined as the mass within $r_{200}$, the radius that encloses a density $\rho(<r_{200}) = 200 \rho_{\rm m}(z)$. The halo mass is a free parameter for both centrals, $M_{\rm cen}$, and satellite galaxies, $M_{\rm sat}$. The concentration normalization $f_{\rm c}$ is only a free parameter for centrals. Since the satellite contribution to the ESD profile is too small to constrain both $M_{\rm sat}$ and $f_{\rm c}^{\rm sat}$, we fix the latter to 1. Apart from the physical concentration of the halo, $f_{\rm c}$ is affected by mis-centering: the off-set between the (assumed) central galaxy and the actual centre of the DM halo. Because $f_{\rm c}$ mitigates the impact of mis-centering, the measured halo mass is not biased by this effect \cite[as shown by][]{viola2015}. In addition to the DM halo, we add the contribution of the stellar component, which is modelled as a point mass with $M = \langle M_* \rangle$, the mean stellar mass of the galaxy sample. This component is added to the contribution of both centrals and satellites.

In the case of the satellite contribution to the ESD profile, the DM halo of the host group is modelled by an offset NFW profile. Each stacked satellite adds a host contribution at its respective projected distance $R_{\rm sat}$ to the group central, such that the total host contribution is integrated over the number distribution $n(R_{\rm sat})$ (see \citealp{sifon2015} for a more detailed description). The two ESD components related to satellite galaxies are multiplied by the satellite fraction $f_{\rm sat}$: the fraction of satellites with respect to the total number of galaxies (including satellites, centrals and non-group galaxies). The ESD component due to centrals (real centrals and non-group galaxies) is in turn multiplied by the central fraction $(= 1-f_{\rm sat})$. The values of the satellite fraction for the (shuffled) cosmic environments are shown in Table \ref{tab:lenses}. As expected, the fraction of satellites increases with the density of the cosmic environment.

At scales above $200 \hk$ the neighbouring host haloes add significant contribution to the ESD signal, known as the 2-halo term. This term is modelled by the two-point matter correlation function $\xi(z, r)$ \cite[]{bosch2002powerspectrum}, which is multiplied by the empirical bias function $b(M)$ \cite[]{tinker2010bias}. As the halo mass $M$ we use the average mass of the central haloes $M_{\rm cen}$. Because we expect that the 2-halo term varies significantly depending on the average density of each cosmic environment, and the correlation function was measured by averaging over all space, we multiply the $\xi(z, r)$ term by a free parameter: the 2-halo amplitude $A_{\rm 2h}$. The final 2-halo contribution becomes:
\begin{equation}
\Delta\Sigma^{\rm 2h}_{\rm cen}(R) = A_{2h} \, b(M_{\rm cen}) \, \Delta\Sigma(\xi(R)) \, ,
\end{equation}
which allows for the flexibility to cover ESD profiles in environments of various densities.

In total, the full model contains four ESD components contributing to the total ESD profile: the central and satellite components ($\Delta\Sigma^{\rm 1h}_{\rm cen}$ and $\Delta\Sigma^{\rm 1h}_{\rm sat}$, modelled by an NFW profile and stellar point mass), the host term corresponding to the satellites ($\Delta\Sigma^{\rm 1h}_{\rm host}$, modelled by an off-set NFW) and the 2-halo term ($\Delta\Sigma^{\rm 2h}_{\rm cen}$, modelled by a scaled matter correlation function):
\begin{multline}
\Delta\Sigma(R) = 
(1-f_{\rm sat}) \times \Delta\Sigma^{\rm 1h}_{\rm cen}(R | M_{\rm cen}, f_{\rm c}^{\rm cen}) \: + \\
\;\;\;\;\;\;\;\;\;\; f_{\rm sat} \times ( \Delta\Sigma^{\rm 1h}_{\rm sat}(R | M_{\rm sat}) + 
\Delta\Sigma^{\rm 1h}_{\rm host}(R | M_{\rm cen}) ) \: + \\
\Delta\Sigma^{\rm 2h}_{\rm cen}(R | A_{\rm 2h}, M_{\rm cen}) \, .
\label{eq:model}
\end{multline}
Together, these four components contain four free parameters: the average halo mass of centrals ($M_{\rm cen}$) and satellites ($M_{\rm sat}$), the concentration parameter of centrals ($f_{\rm c}^{\rm cen}$), and the 2-halo amplitude ($A_{\rm 2h}$), as shown inside the brackets of Eq. (\ref{eq:model}) (where, for brevity, fixed parameters are not shown). Each of these parameters is free for all four cosmic environments, such that our model contains a total of 16 free parameters. The priors of all free parameters are shown in Table \ref{tab:priors}, while all fixed values used in the fit can be found in Table \ref{tab:lenses}.
$\space$

\begin{table*}
	\centering
	\caption{Priors and median posterior values (with 16th and 84th percentile error bars) of the free parameters in the model fit: the average mass of central and satellite galaxies (in units of $[10^{12} \msun$]), the central concentration parameter and the 2-halo amplitude, in both true and shuffled cosmic environments.}
	\renewcommand{\arraystretch}{1.5}
	\begin{tabular}{|lllll|}

\hline
parameter    &    $M_{\rm cen}$    &    $M_{\rm sat}$    &    $f_{\rm c}^{\rm cen}$    &    $A_{\rm 2h}$     \\ \hline
prior type    &    flat    &    flat    &    Gaussian    &    flat     \\ \hline
prior range    &    [0.1, 10]    &    [0.01, 5]    &    $\mu=1, \sigma=0.3$    &    [0, 20]     \\ \hline
\hline
voids    &    $0.75^{+0.31}_{-0.28}$    &    $2.10^{+1.85}_{-1.41}$    &    $1.05^{+0.21}_{-0.19}$    &    $0.93^{+0.72}_{-0.60}$     \\ \hline
sheets    &    $0.74^{+0.63}_{-0.44}$    &    $1.81^{+2.16}_{-1.40}$    &    $0.77^{+0.25}_{-0.24}$    &    $1.50^{+0.79}_{-0.78}$     \\ \hline
filaments    &    $1.43^{+0.86}_{-1.02}$    &    $0.72^{+2.00}_{-0.54}$    &    $0.73^{+0.22}_{-0.22}$    &    $5.22^{+1.18}_{-1.22}$     \\ \hline
knots    &    $0.94^{+1.06}_{-0.63}$    &    $1.00^{+1.15}_{-0.70}$    &    $0.95^{+0.22}_{-0.21}$    &    $10.30^{+2.27}_{-2.25}$     \\ \hline
\hline
shuffled voids    &    $0.77^{+0.41}_{-0.33}$    &    $2.36^{+1.80}_{-1.68}$    &    $0.87^{+0.20}_{-0.21}$    &    $1.81^{+0.79}_{-0.78}$     \\ \hline
shuffled sheets    &    $0.48^{+0.48}_{-0.28}$    &    $2.20^{+1.84}_{-1.64}$    &    $0.88^{+0.24}_{-0.23}$    &    $2.77^{+0.70}_{-0.65}$     \\ \hline
shuffled filaments    &    $1.29^{+0.75}_{-0.90}$    &    $0.72^{+2.01}_{-0.53}$    &    $0.77^{+0.25}_{-0.25}$    &    $4.03^{+0.95}_{-0.96}$     \\ \hline
shuffled knots    &    $1.11^{+1.20}_{-0.71}$    &    $1.81^{+1.51}_{-1.18}$    &    $0.99^{+0.22}_{-0.21}$    &    $9.45^{+1.85}_{-1.79}$     \\ \hline

	\end{tabular}
	\label{tab:priors}
\end{table*}

This model is fitted to the ESD profiles of the four cosmic environments using the {\scshape emcee} sampler \cite[]{foreman2013emcee}, which ingests our model into a Markov Chain Monte Carlo (MCMC). During the fitting procedure, a number of walkers $N_{\rm walkers}$ is moving through the parameter space for a designated number of steps $N_{\rm steps}$, where the direction of each next step is based on the affine invariance method \cite[]{goodman2010mcmc}. Using the Gelman-Rubin convergence diagnostic \cite[]{gelman1992}, we find that we need $N_{\rm walkers} = 100$ and $N_{\rm steps} = 5000$ for our chain to converge. Of the resulting 500,000 evaluations the first 100,000 are discarded as the burn-in phase, leaving a total of 400,000 evaluations. From these evaluations we estimate the values of the free parameters by taking the median (50th percentile), and their $1 \sigma$ uncertainties by taking the 16th and 84th percentile. The minimal $\chi^2$ of our chains is $28.0$ for true cosmic environments, and $34.2$ for shuffled environments. Since the four cosmic environments combined contain $4 \times 10 = 40$ data-points and  $4 \times 4 = 16$ free parameters, the number of degrees of freedom (equal to the expected minimum $\chi^2$) is $N_{\rm dof} = 40-16 = 24$. Consequently the reduced $\chi^2$ of our chains is $28.0/24=1.17$ for true cosmic environments, and $34.2/24=1.43$ for shuffled environments. In the Gaussian case the uncertainty on $\chi^2$ is $\sigma_{\chi^2} = \sqrt{2(N_{\rm dof})} = 6.93$ \cite[]{gould2003chi2}. This indicates that the minimum $\chi^2$ value lies well within $1 \sigma$ of the expected minimum $\chi^2$ for true environments, and just outside for shuffled environments.

\section{Results}
\label{sec:results}

\begin{figure*}
	\includegraphics[width=\textwidth, angle=0]{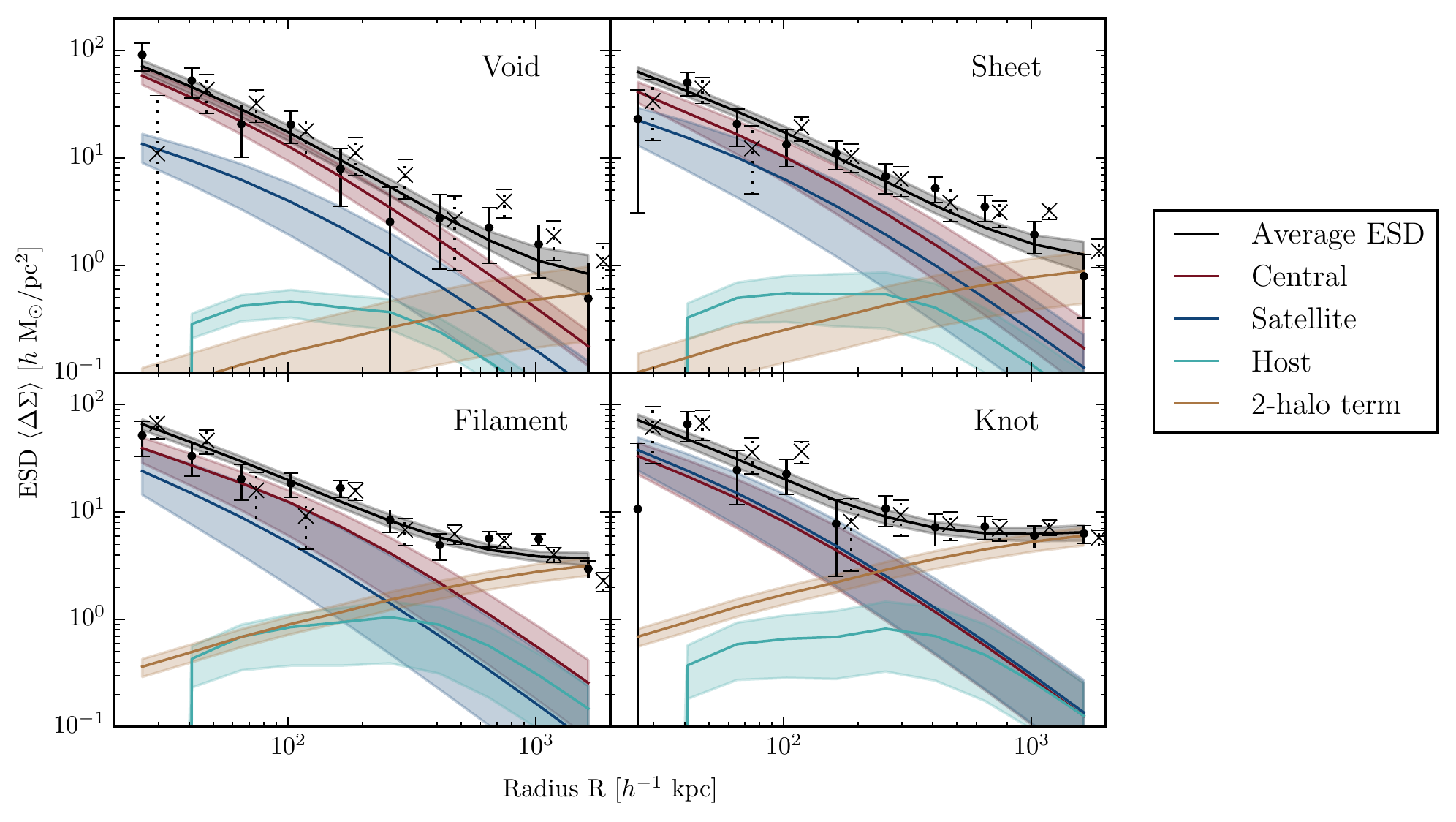}
	\caption{The ESD profiles of GAMA galaxies, stacked according to their cosmic environment and weighted to correct for differences in the stellar mass distribution. The four panels represent the different cosmic environments. The dots with $1\sigma$ error bars represent the measurement, while the lines with error bands show the median and 16th/84th percentile of different components of the ESD model fit to the ESD measurement. The model consists of a central term (red), a satellite term (blue) with corresponding offset host halo term (cyan), and a 2-halo term (yellow). All these terms combine into the total median profile, shown in black. For comparison, we also show the ESD measurement from the shuffled environments as crosses with dotted $1\sigma$ error bars.}
	\label{fig:ESD_envS4}
\end{figure*}

\begin{figure*}
	\includegraphics[width=\textwidth, angle=0]{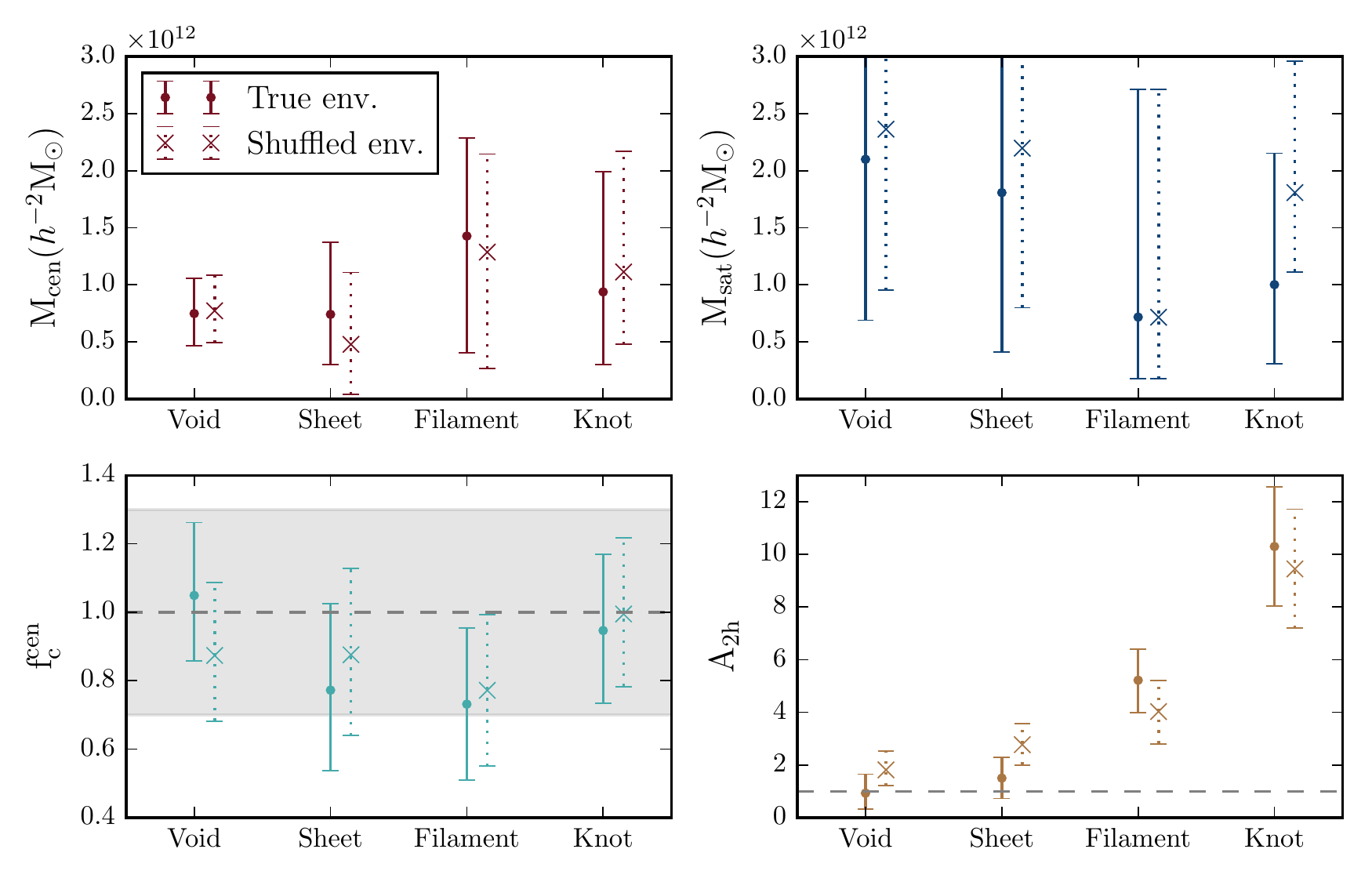}
	\caption{The median parameters resulting from the model fit to the ESD profiles in the four cosmic environments, including 16th and 84th percentile error bars. The top panels show (from left to right) the average halo mass of the centrals (red) and satellite galaxies (blue). The bottom left panel shows the concentration parameter $f_{\rm c}$ of the centrals (cyan), with a band representing the centre $\mu$ and dispersion $\sigma$ of its Gaussian prior. The bottom right panel shows the amplitude of the 2-halo term (yellow), where the dashed line shows the fiducial value. The data points with solid error bars represent the parameters extracted from the ESD profiles of true cosmic environments, while the crosses with dotted error bars represent those from the shuffled environments.}
	\label{fig:ESD_envS4_masses}
\end{figure*}

In order to determine the masses of galaxy haloes as a function of their location in the cosmic web, we measure the average GGL profiles (as detailed in Sect. \ref{sec:lensing}) in each of the four cosmic environments (defined in Sect. \ref{sec:environment}). These ESD profiles are corrected for the measured increase in stellar mass in increasingly dense cosmic environments (shown in Sect. \ref{sec:mstarweight}). By fitting our ESD model (as described in \ref{sec:halomodel}) to these data, we determine the average halo mass of galaxies in each cosmic environment. By applying an identical model to the stacked ESD profiles of galaxies in shuffled cosmic environments (defined in \ref{sec:shuffled}), we can compare their resulting fit parameters. Because the only information from the true cosmic environments that goes into the shuffled environments is that of their local density distribution, any difference between these parameters indicates an effect caused by the cosmic environment alone, i.e. not due to the effects of the local density $\delta_4$. The resulting model fit to the ESD profiles of the four cosmic environments is shown in Fig. \ref{fig:ESD_envS4}. The resulting values of the free parameters: $M_{\rm cen}$, $M_{\rm sat}$, $f_{\rm c}^{\rm cen}$ and $A_{\rm 2h}$, for both the true and the shuffled environments can be found in Table \ref{tab:priors} and Fig. \ref{fig:ESD_envS4_masses}.

In the case of the true cosmic environments, the average halo mass of central galaxies $M_{\rm cen}$ remains constant within the measured uncertainties. On the halo mass of satellite galaxies $M_{\rm sat}$, our model fit provides almost no constraints. When using a flat prior on the concentration of the central halo, we find that the value of $f_{\rm c}^{\rm cen}$ decreases drastically for denser cosmic environments. This is not due to a physical decrease in the concentration of the central halo, as is apparent from the ESD profiles of the galaxy sample containing only centrals. These profiles do not show a significant decrease in the concentration of the central halo in denser cosmic environments (see e.g. the `knot' ESD profile in Fig. \ref{fig:ESD_envS4_ranks}). It is more likely that the increasing signal at larger scales is caused by the increasing number of satellites and neighbouring haloes in denser environments, since the satellite host and 2-halo terms are degenerate with the central concentration. Based on this information we use a Gaussian prior with a central value of $\mu=1$ and a standard deviation of $\sigma = 0.3$, which prevents the decrease of $f_{\rm c}^{\rm cen}$. In the resulting fit, the rising signal at larger scales is accounted for by the satellite host and 2-halo terms. As the density of the cosmic environment increases, $A_{\rm 2h}$ increases by a factor $\sim10$. This behaviour is expected because, due to the increase in the local density $\delta_4$ for increasingly dense environments (see Fig. \ref{fig:envhist}) the contribution of neighbouring haloes to the lensing signal increases.

In Fig. \ref{fig:ESD_envS4_masses} we compare the resulting parameter values with those found using shuffled environments, and find that all parameter values are the same within the $1\sigma$ error bars. Considering the width of the posterior distributions we conclude that there is no measurable difference between the parameters in true and shuffled cosmic environments, suggesting that the dominant effect on halo mass is that of the local density.

\section{Discussion and conclusion}
\label{sec:discon}

We measure the galaxy-galaxy lensing (GGL) signal of 91195 galaxies (within $0.039 < z < 0.263$) from the spectroscopic Galaxy And Mass Assembly (GAMA) survey that overlap with the first $109 \deg^2$ of photometric data from the Kilo-Degree Survey (KiDS). We use the GGL signal to measure the average halo mass of galaxies in four different cosmic environments: voids, sheets, filaments and knots, classified by \cite{eardley2015}. We create a corresponding set of shuffled environments which retain the distribution in local density (the galaxy overdensity within $4 \hm$) of the true environments, but lose the information bound to the cosmic web environment. By comparing the average halo masses from galaxy samples in true and shuffled cosmic environments, we isolate the effect of the cosmic environments on galaxy halo masses from that of the local density. We extract the average halo masses from the measured Excess Surface Density (ESD) profiles by fitting a simple model consisting of a central, a satellite, an off-set host and a 2-halo contribution to the GGL signal. After correcting for the increase in the stellar mass of galaxies in increasingly dense cosmic environments, we find no difference in the average halo mass of central galaxies in different cosmic environments. Our constraints on the average mass of satellite galaxies are to weak to make any statements. The amplitude of the 2-halo term, however, increases significantly from voids to knots. This increase in the 2-halo contribution to the ESD profile is expected, as the local density (within $4\hm $) increases with the density of the cosmic environment.

The posterior distributions of the obtained parameters show no significant difference between the haloes in true and shuffled cosmic environments. We can conclude that, within the statistical limits of our survey, the cosmic environment has no measurable effect on galaxy halo mass apart from the effects related to the local density. This null-result is in agreement with the study of \cite{eardley2015}, who found a strong variation in the Luminosity Function (LF) of galaxies in the four cosmic environments, but no significant difference between the LF in true and shuffled cosmic environments, concluding that the measured effect on the LF could be entirely attributed to the difference in local density of the galaxy populations. Using N-body simulations \cite{alonso2015} studied the dependence of the DM halo mass function on the four cosmic environments. Although they found a strong correlation of the conditional mass function with cosmic environment, they showed that this is caused by the coupling of the cosmic environments to the local density. Using a different classification of GAMA galaxies into cosmic environments (filaments, tendrils and voids), \cite{alpaslan2015environment} measured the effect of the cosmic web on energy output, $u-r$ colour, luminosity, metallicity and morphology of galaxies in both cosmic and local environment. In order to remove the effect due to the difference in the stellar mass distributions, they resampled the galaxy population from each cosmic environment. They found that, as long as they apply this correction, the properties of galaxies in different cosmic environments are approximately identical, and concluded that the effects of large-scale structure on galaxy properties are negligible with respect to the effects from stellar mass and local environment. \cite{darvish2014cosmic}, who measured the star formation rate (SFR) of galaxies as a function of another cosmic environment classification scheme (fields, filaments, and clusters), found that their observed stellar mass and median SFR, as well as the SFR-mass relation and specific SFR, are mostly independent of environment. They did, however, find a significant increase in the fraction of star forming galaxies in filaments. Although the sub-dominance of the effect of large-scale structure on galaxy properties was foreshadowed by many studies, this is the first direct measurement of the effect of the cosmic web on galaxy halo mass.

Based on our results we conclude that, after correcting for local density and stellar mass, the cosmic environments alone  have no measurable effect on DM halo parameters. Even if such an effect exists, future lensing studies would need to reduce the uncertainties on the posteriors found in this study by at least a factor $\sim 3$. Assuming the same approach is used, these studies would require a KiDS-like photometric lensing survey overlapping with a GAMA-like spectroscopic survey of approximately ten times the size of our current $\sim 100 \deg^2$ of overlapping data. Of the present-day photometric lensing surveys, the Dark Energy Survey \cite[DES,][]{des2015} currently has $139 \deg^2$ at its disposal, which is planned to increase to $\sim 5000 \deg^2$ over the next five years. However, DES currently has no overlap with a spectroscopic survey of the area and completeness of the GAMA survey. The Dark Energy Spectroscopic Instrument \cite[DESI,][]{desi2013}, which will serve this purpose, is planned to start nominal operation in 2019. Over the next five years KiDS is planning to observe $\sim 1500 \deg^2$ of the sky \cite[]{dejong2013kids}, overlapping with $\sim 700 \deg^2$ of spectroscopic data from the Wide Area Vista Extragalactic Survey \cite[WAVES,][]{waves2015}, nearing the precision needed to find or rule out an effect from the cosmic web on galaxy haloes. Furthermore, we expect that the Euclid mission \cite[]{euclid2011}, with a planned $\sim 15,000 \deg^2$ of high-quality lensing data, will be able to confirm or negate a possible effect from cosmic environments with very high significance. However, with an estimated launch in 2020 and a nominal mission period of five years, this would take at least five more years to accomplish. Taking all these future missions into consideration we conclude that, using the technique described here, it is unlikely that a direct effect of cosmic environment on halo mass can be measured within the next four to five years.

\section*{Acknowledgements}

We thank Alexander Mead for useful comments on the manuscript. C. Heymans, M. Viola, M. Cacciato, H.Hoekstra, C. Sif\'on, A. Choi  and  C. Heymans acknowledge  support  from  the  European  Research Council  under  FP7  grant  number  279396  (MV,  MC,  CS,  H.Ho), grant number 240185 (AC and CH) and grant number G47112 (CH). M. Viola acknowledges support from  the Netherlands Organisation for Scientific Research (NWO) through grants 614.001.103. H. Hildebrandt is supported by an Emmy Noether grant (No. Hi 1495/2-1) of the Deutsche Forschungsgemeinschaft. R. Nakajima acknowledges support from the German Federal Ministry for Economic Affairs and Energy (BMWi) provided via DLR under project no. 50QE1103. T. M. Roberts and P. Norberg acknowledge support from an European Research Council Starting Grant (DEGAS-259586). This work is supported by the Deutsche Forschungsgemeinschaft in the framework of the TR33 `The Dark Universe'. G. Verdoes Kleijn acknowledges financial support from the Netherlands Research School for Astronomy (NOVA) and Target. Target is supported by Samenwerkingsverband Noord Nederland, European fund for regional development, Dutch Ministry of economic affairs, Pieken in de Delta, Provinces of Groningen and Drenthe. E. van Uitert acknowledges support from an STFC Ernest Rutherford Research Grant, grant reference ST/L00285X/1.

This research is based  on  data  products  from  observations  made  with  ESO Telescopes at the La Silla Paranal Observatory under programme IDs 177.A-3016, 177.A-3017 and 177.A-3018, and on data products produced by Target OmegaCEN, INAF-OACN, INAF-OAPD and  the  KiDS  production  team,  on  behalf  of  the  KiDS  consortium. OmegaCEN and the KiDS production team acknowledge support by NOVA and NWO-M grants. Members of INAF-OAPD and INAF-OACN also acknowledge the support from the Department of Physics \& Astronomy of the University of Padova, and of the Department of Physics of Univ. Federico II (Naples).

GAMA is a joint European-Australasian project based around a spectroscopic campaign using the Anglo-Australian Telescope. The GAMA input catalogue is based on data taken from the Sloan Digital Sky Survey and the UKIRT Infrared Deep Sky Survey. Complementary imaging  of  the  GAMA  regions  is  being  obtained  by  a  number  of  independent survey programs including GALEX MIS, VST KiDS, VISTA VIKING, WISE, Herschel-ATLAS, GMRT and ASKAP providing UV to radio coverage. GAMA is funded by the STFC (UK), the ARC (Australia), the AAO, and the participating institutions. The GAMA website is \url{www.gama-survey.org}.

This work has made use of {\scshape python} (\url{www.python.org}), including the packages {\scshape numpy} (\url{www.numpy.org}), {\scshape scipy} (\url{www.scipy.org}) and {\scshape ipython} \cite[]{perez2007ipython}. Plots have been produced with {\scshape matplotlib} \cite[]{hunter2007matplotlib}.

\emph{Author contributions:} All authors contributed to the development and writing of this paper. The authorship list reflects the lead authors (MB, MC) followed by two alphabetical groups. The first alphabetical group includes those who are key contributors to both the scientific analysis and the data products. The second group covers those who have made a significant contribution either to the data products or to the scientific analysis.




\bibliographystyle{mnras}
\bibliography{biblio}



\bsp	
\label{lastpage}
\end{document}